\documentclass[twocolumn,preprintnumbers,amsmath,amssymb,superscriptaddress,footnote]{revtex4-2}
\usepackage[dvipdfmx]{epsfig}  
\usepackage{bm}
\usepackage{ulem}
\usepackage{amsmath,color}

\def\m@thcombine#1#2{
  \setbox0=\hbox{$#1$}
  \setbox1=\hbox{$#2$}
  \ifdim\wd0>\wd1
    \setbox0=\hbox to\wd1{\hss\box0\hss}
  \else
    \setbox1=\hbox to\wd0{\hss\box1\hss}
  \fi
  \mathop{\vcenter{
    \offinterlineskip\box0\box1}}}
\def\lesim{\m@thcombine<\sim}
\def\gesim{\m@thcombine>\sim}

\begin{document}

\preprint{NITEP 203}

\title{Shell-cluster transition in $^{48}$Ti}

\author{M.~Okada}
\affiliation{Department of Physics, Osaka Metropolitan University, Osaka 558-8585, Japan}

\author{W.~Horiuchi}
 \email{whoriuchi@omu.ac.jp}
\affiliation{Department of Physics, Osaka Metropolitan University, Osaka 558-8585, Japan}
\affiliation{Nambu Yoichiro Institute of Theoretical and Experimental Physics (NITEP), Osaka Metropolitan University, Osaka 558-8585, Japan}
\affiliation{RIKEN Nishina Center, Wako 351-0198, Japan}
\affiliation{Department of Physics,
  Hokkaido University, Sapporo 060-0810, Japan}

\author{N.~Itagaki}
 \email{itagaki@omu.ac.jp}
\affiliation{Department of Physics, Osaka Metropolitan University, Osaka 558-8585, Japan}
\affiliation{Nambu Yoichiro Institute of Theoretical and Experimental Physics (NITEP), Osaka Metropolitan University, Osaka 558-8585, Japan}
\affiliation{RIKEN Nishina Center, Wako 351-0198, Japan}

\begin{abstract}
\noindent {\bf Background:}
Whether or not the $\alpha$ ($^4$He nucleus) clustering exists in the medium-mass region of nuclear systems is a fundamental and intriguing question.
However, the recent analysis of the $\alpha$ knockout reaction on $^{48}$Ti [Phys. Rev. C {\bf 103}, L031305 (2021)] poses a puzzle: The microscopic wave function gives an $\alpha$ knockout cross section that is two orders of magnitude smaller than the experiment, while basic nuclear properties such as the charge radius and the electromagnetic transition probabilities are well explained.\\
\noindent{\bf Purpose:}
The ground-state structure of $^{48}$Ti is investigated by using
proton- and $\alpha$-nucleus elastic scattering
at a few to several hundred MeV,
which offers different sensitivity to the region of the nuclear density profiles.\\
\noindent{\bf Method:}
Four types of the density distributions,
the $jj$-coupling shell model and three cluster model configurations,
are generated in a single scheme by
the antisymmetrized quasi-cluster model (AQCM).
The angular distribution of the proton- and $\alpha$-$^{48}$Ti
elastic scattering cross sections are obtained with a reliable high-energy
reaction theory, the Glauber model.\\
\noindent{\bf Results:} 
The $jj$-coupling shell model configuration is found to best reproduce
the proton-nucleus elastic scattering cross section.
On the other hand, the trace of the $\alpha$ cluster structure
in the tail region of the wave function is embedded
in the $\alpha$-nucleus elastic scattering cross section.
\\
\noindent{\bf Conclusions:}
Our results suggest that the structure of the nucleus changes
as a function of distance from the center, from the $jj$-coupling
shell model structure in the surface region
to the $\alpha$+$^{44}$Ca cluster structure in the tail region.
This picture is consistent
with the finding of the $\alpha$ knockout reaction on $^{48}$Ti. 
\end{abstract}
\maketitle

\section{Introduction}

It is well known that $\alpha$ clustering plays a crucial role in light nuclei. 
In addition to the light-mass region,
whether or not the $\alpha$ clustering exists
in the medium-mass region is a fundamental and intriguing question.
However, the degree of the clustering is expected to be smaller because
the effect of the spin-orbit interaction, which acts to break up the $\alpha$ clusters near the nuclear surface and induces the independent nucleon motion
of the $jj$ coupling shell model, becomes stronger with increasing mass number, more precisely 
with increasing the total angular momentum $j$ of single particles~\cite{Mayer}.
A possible candidate for the medium-heavy nucleus
with the cluster structure is  $^{44}$Ti, 
which is a $Z=N$ nucleus. The presence of an $\alpha+^{40}$Ca structure 
was predicted in Ref.~\cite{Michel86},
and subsequently the inversion doublet structure
was experimentally confirmed~\cite{Yamaya90,Yamaya98}, providing
supporting evidence for the presence of an asymmetric cluster structure.
However, the general persistence of the $\alpha$ cluster structure in the Ti isotopes, including the $\beta$ stable ones with more neutrons in the $pf$-shell, requires further discussion. 

In this respect, the recent analysis of the $\alpha$ knockout reaction
on $^{48}$Ti poses a vexing puzzle.
It is presumed that $^{48}$Ti has less $\alpha$ cluster components
compared to $^{44}$Ti and $^{52}$Ti~\cite{Bailey19}; nevertheless,
the $\alpha$ particle is knocked out with a certain cross section.
However, the wave function obtained with the structure calculation based on the antisymmetrized molecular dynamics (AMD) exhibits the dominance of
the mean-field type and
gives the $\alpha$ knockout reaction cross section that is two orders of magnitude smaller than the experimental one~\cite{Taniguchi21}.
The cross section can indeed be explained if the presence of an $\alpha+^{44}$Ca cluster structure is assumed with a huge relative distance of about 4.5~fm but other fundamental properties of $^{48}$Ti such as the charge radius and the electromagnetic transition probabilities cannot be explained with this cluster wave function.

With the aim of providing some insight into this question,
we study the ground state of $^{48}$Ti.
In fact, the cross section of the $\alpha$ knockout reaction is only sensitive to the $\alpha$ clustering in the tail region of the wave function.  
This is because the information about the more inner region of the wave function is drowned out by the strong $\alpha$ absorption. The transition matrix density shows 
that the $\alpha$ knockout reaction tells us nothing about the character of the wave function within the radius of 5~fm~\cite{Taniguchi21}.
Therefore, even if the more inner part of the wave function is different from a simple $\alpha$ cluster structure, it does not affect the $\alpha$ knockout cross section.

In this paper, we use proton and $\alpha$ particles
to probe the ground state properties of $^{48}$Ti.
The medium- to high-energy elastic scattering
is useful to study the nuclear density profiles,
enabling one to distinguish whether $^{48}$Ti is $\alpha$ cluster-like
or $jj$-coupling shell-like in this mass region.
Here, both the wave functions for the shell and cluster configurations 
are consistently produced in a single scheme, which is achieved by
using the antisymmetrized quasi-cluster model (AQCM)~\cite{AQCM01,AQCM02,AQCM03,AQCM04,AQCM05,AQCM06,AQCM07,AQCM08,AQCM09,AQCM10,AQCM11,AQCM12,AQCM13,AQCM14}. This model allows us to smoothly transform the cluster model wave function into the $jj$-coupling shell model wave function, and we can treat the two on the same footing. 
The analyses of the proton-nucleus elastic scattering
for the ground states of $^{44}$Ti and $^{52}$Ti
were already carried out by combining AQCM
and the Glauber model~\cite{AQCM-Glauber-1},
showing significant difference in the cross sections, especially,
at around the first diffraction peak.
Unfortunately, no experimental result for $^{44}$Ti and $^{52}$Ti is available.
In the present case of $^{48}$Ti, which is a $\beta$-stable Ti isotope,
there are experimental data to be compared.
Similar studies of the distinction between
the cluster and shell densities have been carried out for 
light nuclei such as $^{12}$C, $^{16}$O~\cite{AQCM-Glauber-2},
and $^{20}$Ne~\cite{AQCM-Glauber-3}.

The paper is organized as follows.
Section~\ref{methods.sec} summarizes the formulation of the present approach.
We briefly explain how to calculate the density distributions
for shell and cluster configurations using
the AQCM and the elastic scattering cross section
with a high-energy reaction theory, the Glauber model.
Our results are presented in Sec.~\ref{results.sec}.
We discuss the relationship between density profiles
of the model wave functions 
and observables such as the proton- and $\alpha$-nucleus
elastic scattering cross sections.
Finally, the conclusion is given in Sec.~\ref{conclusion.sec}.

\section{Methods}
\label{methods.sec}

\subsection{Shell model type and cluster model type wave functions based on AQCM}
\label{AQCM.sec}

Based on AQCM, we introduce shell model type (S-type) and cluster model type (C-type) wave functions.
In both cases, the single-particle wave function has a Gaussian shape as in the Brink model~\cite{Brink}
\begin{align}
  \phi_i=  \left(\frac{2\nu}{\pi}\right)^{3/4}\exp\left[-\nu(\bm{r}_i-\bm{\zeta}_{i})^2\right]\chi_{i}\eta_{i},
\label{spwf}
\end{align}
where $\chi_i$ and
$\eta_i$ are the spin and the isospin parts of the wave function, respectively.
The parameter $\nu$ is a size parameter, and $\bm{\zeta}_i$ is the Gaussian center parameter.
The total wave function $\Phi$ is the antisymmetrized product
of these single-particle wave functions
\begin{align}
  \Phi=\mathcal{A}
  \left\{\prod_{i=1}^{A} \phi_i \right\},
\end{align}
where $\mathcal{A}$ is the antisymmetrizer
and $A=48$ is the mass number.

These forty-eight single-particle wave functions consist of a $^{40}$Ca core and eight valence nucleons.
The $^{40}$Ca core can be described as ten $\alpha$ clusters with small relative distances, which is consistent with the shell model description of $^{40}$Ca due to the antisymmetrization effect.
Each $\alpha$ cluster is defined as four nucleons (proton spin-up, proton spin-down, neutron spin-up, neutron spin-down) sharing a common value for the Gaussian center parameter $\bm{\zeta}_{i}$.
The actual positions of the ten $\alpha$ clusters for the $^{40}$Ca core are described in Ref.~\cite{AQCM06}.

For the eight valence nucleons, we first introduce three $\alpha$ clusters ($^{12}$C) around the $^{40}$Ca core and then remove four protons afterwards. These three $\alpha$ clusters are introduced to have an equilateral triangular shape with a small relative distance
around the $^{40}$Ca core.
The spin parts of the single particles in the three $\alpha$ clusters are also introduced with the equilateral triangular symmetry as described in Ref.~\cite{AQCM06}.
These single-particle orbits in the three $\alpha$ clusters are excited to the $pf$-shell due to the antisymmetrization effect with the nucleons in the $^{40}$Ca core.
However, there is no spin-orbit contribution yet unless the $\alpha$ clusters are broken.
Therefore, next,
these single-particle orbits are transformed into the $f_{7/2}$ orbits of the $jj$-coupling shell model
by giving the imaginary parts to the Gaussian center parameters
based on the transformation of AQCM~\cite{AQCM05}
\begin{align}
  \bm{\zeta}_i=\bm{R}_i+i\Lambda \bm{e}_i^{\rm spin}\times \bm{R}_i.
\label{AQCM}
\end{align}
Here $\bm{R}_i$ represents the spatial location of the $i$th single particle,
and $\bm{e}^{\rm spin}_i$ is a unit vector for the intrinsic spin.
The imaginary parts of the Gaussian center parameters represent imparted momenta to the nucleons,
and $\alpha$ clusters are broken in such a way that spin-up and spin-down nucleons are boosted in opposite directions and perform time-reversal motions.
The parameter $\Lambda$ controls the breaking of the $\alpha$ clusters,
and the $^{48}$Ti wave function is constructed by removing four protons from the twelve nucleons around the $^{40}$Ca core.

For the shell model wave function, S-type, to break clusters,
the $\Lambda$ value is set to 1 for all the eight nucleons.
In this way,
the $jj$-coupling shell model wave function of $^{48}$Ti [$(f_{7/2})^2$ for the protons and $(f_{7/2})^{6}$ for the neutrons around the $^{40}$Ca core] is generated.

Next, we introduce the cluster model wave function, C-type. 
For the four nucleons (proton spin-up, proton spin-down, neutron spin-up, neutron spin-down) in 
the $f_{7/2}$ orbits around the $^{40}$Ca core,
we set $\Lambda = 0$ in Eq.~(\ref{AQCM})
and remove the imaginary part of the Gaussian center parameters; they are returned 
to an $\alpha$ cluster. This $\alpha$ cluster is separated from the rest ($^{44}$Ca) by the distance of $d$~fm.
After setting all these Gaussian center parameters of S-type and C-type, the whole system is moved to satisfy the condition of $\sum_{i=1}^{48} \left<\bm{r}_i\right>=0$.
\par
Once the model wave function $\Phi$ 
is set, the intrinsic density distribution $\tilde{\rho_t}(\bm{r})$ is obtained by calculating the expectation value of
$ \sum_{i \in t} \delta(\bm{r}_i-\bm{r})$,
\begin{align}
  \tilde{\rho_t}(\bm{r})
  = \langle \Phi |\sum_{i \in t} \delta(\bm{r}_i-\bm{r}) | \Phi \rangle / \langle \Phi | \Phi 
 \rangle,
\end{align}
where the summation is taken over protons ($t=p$) or neutrons ($t=n$).
The center-of-mass wave function can be eliminated by
using a Fourier transform~\cite{Horiuchi07} as
\begin{align}
  \int d\bm{r}\,e^{i\bm{k}\cdot\bm{r}}\rho^{\rm int}_t(\bm{r})=\exp\left(\frac{k^2}{8A\nu}\right)
  \int d\bm{r}\,e^{i\bm{k}\cdot\bm{r}} \tilde{\rho_t}(\bm{r}),
\end{align}
and we use $\rho^{\rm int}_t(\bm{r})$ as the intrinsic density free of center of mass motion.
The density distribution in the laboratory frame
is finally obtained by averaging the
intrinsic density distribution over the angles~\cite{Horiuchi12} as
\begin{align}
  \rho_t(r)=\frac{1}{4\pi}\,\int d\hat{\bm{r}}\,\rho^{\rm int}_t(\bm{r}).
\label{density.eq}
\end{align}

\subsection{Elastic scattering cross section within the Glauber model}
\label{Glauber.sec}

Proton-nucleus elastic scattering
is one of the most direct ways of obtaining information
on the density profile. We remark that 
the full density distribution can be obtained
by measurements up to backward angles~\cite{Terashima08,Zenihiro10}.
As long as the nuclear surface density is of interest,
only the cross sections at the forward angles, i.e.,
the cross section at the first peak in the proton-nucleus
diffraction is needed to extract the ``diffuseness'' of the
density distribution as prescribed in Ref.~\cite{Hatakeyama18}.
To relate the density profile to the reaction observables 
we employ a high-energy microscopic reaction theory,
the Glauber model~\cite{Glauber}.

The differential cross section of the elastic scattering is given by
\begin{align}
  \frac{d\sigma}{d\Omega}=|f(\theta)|^2
\end{align}
with the scattering amplitude of the nucleus-nucleus
elastic scattering~\cite{Suzuki03}
\begin{align}
  f(\theta)=F_C(\theta)+\frac{ik}{2\pi}\int d\bm{b}\, e^{-i\bm{q}\cdot\bm{b}+2i\eta \ln(kb)}\left(1-e^{i\chi_{xT}(\bm{b})}\right),
\label{scat_amp.eq}
\end{align}
where $F_C(\theta)$ is the Rutherford scattering amplitude,
$\bm{b}$ is the impact parameter vector,
and $\eta$ is the Sommerfeld parameter.
As the nuclear scattering occurs in several hundred MeV,
relativistic kinematics is used for the wave number $k$.

Here, we treat proton- or $\alpha$-target nucleus
($xT$; $x=p$ or $\alpha$, T: target nucleus) system.
The optical phase shift function $\chi_{xT}$ contains
all the dynamical information for the $xT$ system
within the Glauber model, but its evaluation involves
multiple integration. For practical calculations,
the optical limit approximation (OLA)~\cite{Glauber,Suzuki03}
is used to compute the optical phase shift function as
\begin{align}
  i\chi_{pT}(\bm{b})\approx -\int d\bm{r}\,
  \left[\rho_p(\bm{r})\Gamma_{pp}(\bm{b}-\bm{s})
  +\rho_n(\bm{r})\Gamma_{pn}(\bm{b}-\bm{s})\right],
\end{align}
for a proton-nucleus system, where a single-particle coordinate is
expressed by $\bm{r}=(\bm{s},z)$ with $z$ being the beam direction.
Further, we evaluate the optical phase shift function for an $\alpha$-nucleus system with
\begin{align}
  i\chi_{\alpha T}(\bm{b})&\approx -\iint d\bm{r}\,d\bm{r}^\prime\,
  \left[\rho^{\alpha}_p(\bm{r}^\prime)\rho_p(\bm{r})
    \Gamma_{pp}(\bm{b}+\bm{s}^\prime-\bm{s})\right.\notag\\
    &+\rho^{\alpha}_p(\bm{r}^\prime)
    \rho_n(\bm{r})\Gamma_{pn}(\bm{b}+\bm{s}^\prime-\bm{s})\notag\\
    &+\rho^{\alpha}_n(\bm{r}^\prime)\rho_p(\bm{r})
    \Gamma_{np}(\bm{b}+\bm{s}^\prime-\bm{s})\notag\\
    &+\left.\rho^{\alpha}_n(\bm{r}^\prime)
    \rho_n(\bm{r})\Gamma_{nn}(\bm{b}+\bm{s}^\prime-\bm{s})
    \right],
\end{align}
where $\rho^\alpha$ is the intrinsic density distribution of
the $\alpha$ particle with the $(0s)^4$ harmonic oscillator configuration and a size parameter 
reproducing the measured charge radius.
The parameterization of the proton-proton (neutron-proton) profile function
$\Gamma_{pp}=\Gamma_{nn}$ ($\Gamma_{pn}=\Gamma_{np}$)
is given in Ref.~\cite{Ibrahim08}.
Once the above inputs are set, the theory has no tunable parameter,
and so, the resulting reaction observables are a direct reflection of
the density profiles of the target nucleus.
This model works well as shown, for example,
in Refs.~\cite{Horiuchi16, Hatakeyama19},
and its accuracy compared to those obtained by the full evaluation
of the optical phase shift function
was discussed in Refs.~\cite{Varga02,Ibrahim09,Nagahisa18,Hatakeyama19}.

\section{Results and discussions}
\label{results.sec}

\subsection{Properties of the wave functions}
\label{wf.sec}

Here we generate the shell-model-like (S-type)
and cluster-model-like (C-type) configurations and show their properties.
The C-type is further subdivided into C-type-1, C-type-2, and C-type-3
depending on the size parameter $\nu$ in Eq.~(\ref{spwf})
and the $\alpha$--$^{44}$Ca distance $d$.
All these configurations reproduce the experimental charge radius of $^{48}$Ti~\cite{Angeli13}.

\begin{table*}[htb]
  \begin{center}
    \caption{Properties of the shell-model-like (S-type)
      and $\alpha$-cluster-model-like (C-type-1, C-type-2, and C-type-3) configurations for $^{48}$Ti. See text for details. The experimental point-proton rms radius ($r_p$) of $^{48}$Ti is 3.50 fm extracted from Ref.~\cite{Angeli13}.}
\begin{tabular}{lcccccccccccc}
\hline\hline
  &&$\nu$ (fm$^{-2}$) &$d$ (fm)&$\left<N\right>$ &$\left<LS\right>$ ($\hbar^2$) &$r_p$ (fm) &$r_n$ (fm) & $r_m$ (fm)&$a_p$ (fm) & $a_n$ (fm) & $a_m$ (fm)\\
\hline
S-type    &&0.1269 &--   &84.0 &12.0 & 3.50 & 3.61 & 3.56 & 0.544&0.531&0.536\\
C-type-1&&0.1267 & 0.1   &84.0 &5.98 & 3.50 & 3.62 & 3.56  & 0.596&0.577&0.586\\
C-type-2&&0.1299 &2.379 &84.6 &5.99 & 3.50 & 3.61 & 3.56   & 0.611&0.585&0.599\\
C-type-3&&0.1395 &4.5     &87.4 &6.00 & 3.50 & 3.58 & 3.54 & 0.636&0.603&0.620\\
\hline\hline
\end{tabular}  
\label{results.tab}
\end{center}
\end{table*}

\subsubsection{Shell-model-like configuration (S-type)}

As shown in Table~\ref{results.tab},
the shell-model-like configuration (S-type)
has a total harmonic oscillator quanta $\left<N\right>$ of 84.0.
This means that eight nucleons with $N=3$ are located outside 
the $^{40}$Ca core with $N=60$.
Given the $\Lambda$ value of 1 in Eq.~(\ref{AQCM}) for the eight nucleons around $^{40}$Ca, the $jj$-coupling shell model state is realized, with two protons and six neutrons occupying the $f_{7/2}$ orbits. 
This is confirmed by the calculation of
the expectation values of the one-body spin-orbit operator
$\sum_{i=1}^{48}\bm{l}_i\cdot\bm{s}_i$, which is listed in the $\left<LS\right>$ column, 
where $\bm{l}_i$ and $\bm{s}_i$ stand for the orbital angular momentum and spin operators
of the $i$th nucleon, respectively. 
Here, there is no contribution from the $^{40}$Ca core part, and one nucleon in 
the $f_{7/2}$ orbit has the $\bm{l}\cdot\bm{s}$ value of 1.5~$\hbar^2$, and thus,
$1.5~\hbar^2 \times 8 = 12~\hbar^2$ is the ideal value of the $jj$-coupling shell model.
We can confirm that our model reproduces this ideal value.
The size parameter $\nu$ of the single particle wave functions in Eq.~(\ref{spwf})
is chosen to be 0.1269~fm$^{-2}$, which reproduces the root-mean-square (rms)
radius of the point-proton (column $r_p$) derived as 3.50~fm.
The rms radii of the point-neutrons ($r_n$) and matter distribution ($r_m$) are obtained as 3.61~fm and 3.56~fm,
respectively.

\subsubsection{$\alpha$-cluster-like configuration (C-type)}

The $\alpha$-cluster-like configuration (C-type) has the structure
of  $^{44}$Ca plus $\alpha$.
This can be obtained by setting $\Lambda = 0$ for the two protons and two neutrons
around $^{44}$Ca. These four nucleons form an $\alpha$ cluster.
Furthermore, the center of this $\alpha$ cluster can be separated from $^{44}$Ca with the $jj$-coupling shell model configuration by the distance of $d$~fm.

We prepare three $\alpha$-cluster-like configurations (C-type-1, C-type-2, and C-type-3).
They have different $\nu$ and $d$ values but all three configurations reproduce the experimentally observed $r_p$.

For C-type-1, the size parameter $\nu$ (0.1267~fm$^{-2}$)
is set so as to reproduce $r_p$ of the subsystem, $^{44}$Ca, 3.42~fm.
Meanwhile, the parameter $d$ for the relative distance between $^{44}$Ca and $^4$He is 
determined to reproduce $r_p$ of the whole system, $^{48}$Ti, 3.50~fm.
As shown in Table~\ref{results.tab}, the $d$ parameter must be very small in this case,
and the resulting configuration is almost like a zero distance limit between $^{44}$Ca and $^4$He.

For C-type-2, the size parameter $\nu$ (0.1299~fm$^{-2}$)
is set so as to reproduce $r_p$ of $^{40}$Ca, 3.38~fm.
Again, the parameter $d$ for the relative distance between $^{44}$Ca and $\alpha$
is determined to reproduce $r_p$ of the whole system, $^{48}$Ti,
and in this case, the finite value of $d = 2.379$~fm is obtained. 

For C-type-3, we mimic the wave function that reproduces the $\alpha$ knockout reaction cross section.
As mentioned above, the $\alpha$ knockout reaction is reproduced by the cluster wave function with the
 $\alpha$--$^{44}$Ca distance of 4.5~fm~\cite{Taniguchi21}. Therefore, here we set $d = 4.5$~fm.
To reproduce $r_p$ of $^{48}$Ti, $\nu=0.1395$~fm$^{-2}$ is required.

As shown in Table~\ref{results.tab}, the harmonic oscillator quanta
$\left<N\right>$ increases with the value of $d$; the $\left<N\right>$ value of 84.0 is obtained for C-type-1 ($d=0.1 $~fm), which increase to 84.6 for C-type-2 ($d=2.379$~fm) and 87.4 for C-type-3 ($d=4.5$~fm).
For the expectation values of the one-body spin-orbit operator, $\left<LS\right>$, 
since the $^{40}$Ca core and $\alpha$ cluster parts do not contribute, the value comes only from the four neutrons
around $^{40}$Ca. The ideal value is 6~$\hbar^2$; a neutron in $f_{7/2}$ has a contribution of 1.5~$\hbar^2$,
and the actual values are close to it, as shown in the column $\left<LS\right>$.

\subsection{Density distributions}
\label{density.sec}

\begin{figure}[ht]
\begin{center}
  \epsfig{file=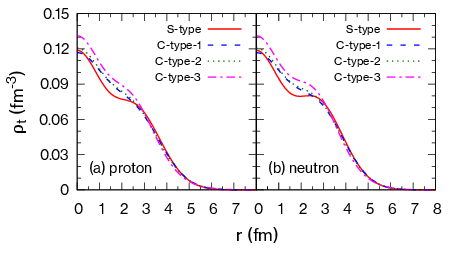, width=\columnwidth}
  \caption{
    Density distributions of $^{48}$Ti for (a) point-protons
    and (b) point-neutrons
    as a function of $r$, the distance from the origin.}
    \label{dens48Ti.fig}
  \end{center}
\end{figure}
\par
Figure~\ref{dens48Ti.fig}~(a) and (b) show the point-proton and point-neutron density distributions of $^{48}$Ti, respectively, as a function of $r$, the distance from the origin.
Despite the fact that all these density distributions give the same charge radii,
they have different density profiles.
In the following subsection, we will discuss the proton-$^{48}$Ti elastic scattering, which is sensitive to the density in the region that
the value is about half of the central one.
This half-density region corresponds
to $r \approx 3$~fm, which we call the surface region.
 Meanwhile, the $\alpha$ knockout reaction is sensitive to the wave function 
$r \gesim 5$~fm~\cite{Taniguchi21}, which can be called the tail region.
 To quantify the density profiles at around the half-density region,
 it is convenient to evaluate the nuclear diffuseness
for proton ($a_p$), neutron ($a_n$), and matter ($a_m$) density distributions
~\cite{Hatakeyama18} by minimizing
\begin{align}
  \int_0^\infty dr\, r^2  |\rho_t(r)-\rho^{\rm 2pF}_t(r)| 
\end{align}
with
\begin{align}
\rho^{\rm 2pF}_t(r)=\frac{\rho_0}{1+\exp\left(\frac{r-R_t}{a_t}\right)}.
\end{align}
Table~\ref{results.tab} lists those calculated diffuseness values. The S-type configuration exhibits the smallest diffuseness values, i.e.,
the sharpest nuclear surface,
and the nuclear surface becomes more diffused as the $\alpha$--$^{44}$Ca cluster
structure develops. As we will see later,
differences in these diffuseness values are actually reflected
in the elastic scattering cross sections.
 
\begin{figure}[ht]
\begin{center}
  \epsfig{file=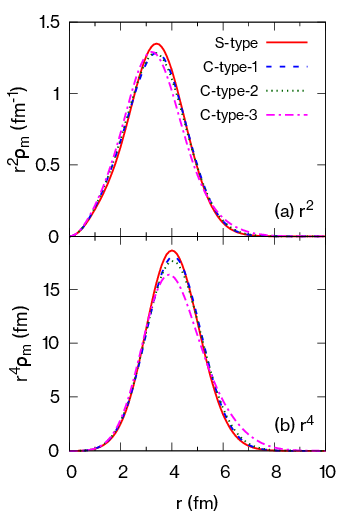, scale=1.2}
  \caption{Matter density distributions of $^{48}$Ti ($\rho_m$)
    multiplied by (a) $r^2$ and (b) $r^4$ as a function of the distance from the origin, $r$.}
    \label{matter-dens.fig}
  \end{center}
\end{figure}

The difference in the density distributions becomes more visible
when $r^{2n}$ ($n$: integer) is multiplied.
Figure~\ref{matter-dens.fig} shows the matter density distributions
of $^{48}$Ti ($\rho_m$) multiplied by (a) $r^2$ and (b) $r^4$.
The integration of $4\pi r^2 \rho_m$ over $r$ gives the particle number,
and thus, in Fig.~\ref{matter-dens.fig}~(a), the four areas that the four lines create together with the horizontal axis are equal.
We see some differences beyond the half-density radius,
$r \gesim 3$~fm in the $r^2\rho_m$ distribution.
In Fig.~\ref{matter-dens.fig} (b),
the distribution of C-type-3 is significantly shifted
to the larger $r$ side compared to the other three lines.
This characteristic feature of C-type-3 shown in the $r^4 \rho_m$
distribution stems from its large clustering,
which affects the $\alpha$-nucleus elastic scattering cross section
at the first diffraction peak, where the signature of the large $\alpha$ clustering is embedded, as we will discuss later.

\subsection{Proton-$^{48}$Ti and $\alpha$-$^{48}$Ti elastic scattering}

\begin{figure}[ht]
\begin{center}
  \epsfig{file=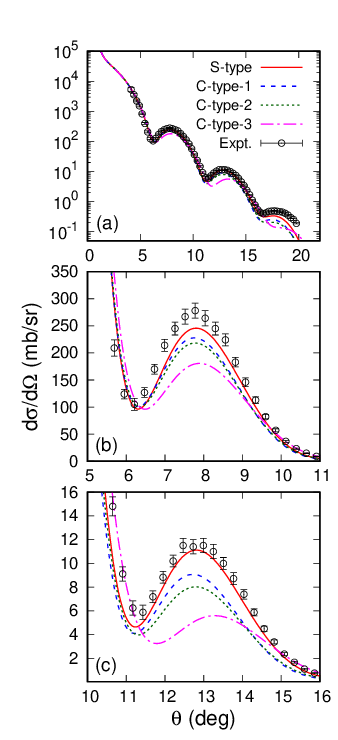, scale=1.3}                    
  \caption{
  Differential cross sections for the proton-$^{48}$Ti elastic scattering 
  at the incident energy of 1000~MeV 
in logarithmic (a) and linear (b), (c) scales
as a function of the scattering angle. See text for details.
The experimental data (incident energy of 1044~MeV)
are taken from Ref.~\cite{Alkhazov76}.
 }
    \label{dcs1000.fig}
  \end{center}
\end{figure}

These differences between the density distributions are reflected in the diffraction patterns of the proton-nucleus  elastic scattering.
Figure~\ref{dcs1000.fig} shows the differential cross section for the proton-$^{48}$Ti
elastic scattering.
The proton incident energy is chosen to be 1000~MeV,
and the experimental data (incident energy of 1044~MeV)
are taken from Ref.~\cite{Alkhazov76}.
Three of the four nuclear densities, except for C-type-3, are in reasonable agreement with the experimental result up to the second peak [Fig.~\ref{dcs1000.fig}~(a)], but for a more accurate comparison,
we plot the cross sections in a linear scale in Fig.~\ref{dcs1000.fig}~(b) and (c).
The angle and height of the first peak position reflect the size and diffuseness
of the target nucleus,
and Fig~\ref{dcs1000.fig}~(b) shows the experimental data around this region is best reproduced by the S-type configuration. 
This means that the $^{48}$Ti nucleus has a density distribution close to the $jj$-coupling 
shell model picture around its surface region.
Figure~\ref{dcs1000.fig}~(c) shows the cross sections around the second peak,
and the agreement between the experimental data and the result of the S-type becomes even better than for the first peak region.

\begin{figure}[ht]
\begin{center}
  \epsfig{file=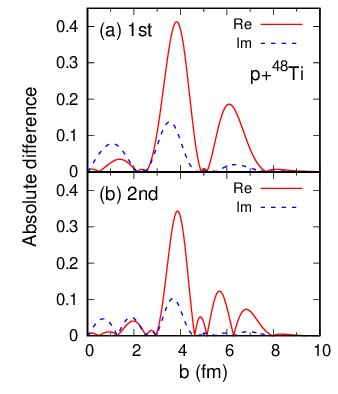, scale=1.4}                    
  \caption{
    Absolute difference of the real and imaginary parts of
    the radial scattering amplitudes [Eq.~(\ref{g.eq})]
    at the (a) first and (b) second diffraction peaks
    of the S-type and C-type-3 configurations for proton-$^{48}$Ti scattering
    at the incident proton energy of 1000 MeV.
    See text for details.
 }
    \label{ampdiff-p.fig}
  \end{center}
\end{figure}

What regions of the density profiles are actually observed?
To answer this question, it is intuitive to look at the radial dependence
of the scattering amplitude, i.e.,
integrand of the second term of Eq.~(\ref{scat_amp.eq}) at
a specific scattering angle~\cite{Hatakeyama18},
which is explicitly written by
\begin{align}
  g(\theta,b)=ikb\,e^{-2ikb\sin\left(\frac{\theta}{2}\right)+2i\eta\ln(kb)}
    \left(1-e^{i\chi_{xT}(b)}\right).
\label{g.eq}
\end{align}
It is worthwhile to recall the relation
\begin{align}
f(\theta)=F_C(\theta)+\int_{0}^\infty db\, g(\theta,b).
\end{align}
We set $\theta$ near the first and second diffraction peaks
and compare $g$ with the different AQCM configurations
as a function of the impact parameter $b$.
Here we take S-type and C-type-3 configurations,
where the most different results are expected.

Figure~\ref{ampdiff-p.fig} displays
the absolute difference of the real and imaginary parts
of $g$ between S-type and C-type-3 configurations
for the differential elastic scattering cross sections
at around (a) the first peak ($\theta=7.8^\circ$)
and (b) the second peak ($\theta=13^\circ$) positions.
At the first peak position [Fig.~\ref{ampdiff-p.fig}~(a)], the absolute difference in $g$ is largest at the surface region $b\approx 4$~fm,
which is consistent with our basis that the nuclear diffuseness is most reflected in
the first diffraction peak~\cite{Hatakeyama18}, considering that the range of the nucleon-nucleon interaction is about 1~fm.
At the second peak position [Fig.~\ref{ampdiff-p.fig}~(b)], 
the difference is also largest at the surface region. 
The second peak region includes
the information of the density wider than the surface region,
and there the picture of the $jj$-coupling shell model works well.
Here, the contribution around the tail region ($b \approx 6$~fm) in Fig.~\ref{ampdiff-p.fig}~(b) 
is reduced from that in Fig.~\ref{ampdiff-p.fig}~(a) compared to the reduction of the surface region ($b \approx 4$~fm),
and thus, the second peak reflects the difference of S-type and C-type-3 around the surface region more pronouncedly than the first peak.

From the analysis of the proton-nucleus elastic scattering, we have found that
the $^{48}$Ti nucleus has a $jj$-coupling shell model structure rather than the $\alpha+^{44}$Ca cluster structure. 
However, as discussed earlier, the $\alpha$ knockout reaction cross section, 
which is sensitive to the tail region of the wave function,
is explained by the $\alpha+^{44}$Ca cluster structure
with a large relative distance.

\begin{figure}[ht]
\begin{center}
  \epsfig{file=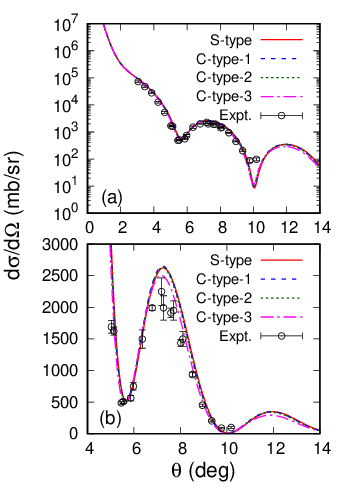, scale=1.3}                    
  \caption{
  Differential cross sections for the $\alpha+^{48}$Ti elastic scattering 
at the incident $\alpha$ particle energy of 240~MeV 
in logarithmic (a) and linear (b) scales as a function of the scattering angle.
 The experimental data are taken from Ref~\cite{Tokimoto2006}.
 }
    \label{dcs-alpha.fig}
  \end{center}
\end{figure}

\begin{figure}[ht]
\begin{center}
  \epsfig{file=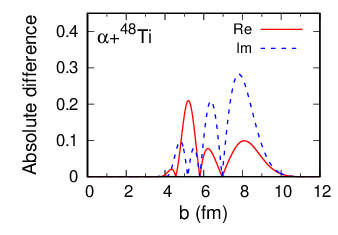, scale=1.4}                    
  \caption{
    Same as Fig.~\ref{ampdiff-p.fig} but at around
    the first peak position of the 
    $\alpha+^{48}$Ti elastic scattering cross section at
    the incident $\alpha$ particle energy of 240 MeV.
 }
    \label{ampdiff-a.fig}
  \end{center}
\end{figure}

Indeed, we can deduce the vestige of the $\alpha$ clustering in the tail region
of the wave function from the $\alpha$-nucleus elastic scattering cross section.
Figure~\ref{dcs-alpha.fig} shows the differential cross section
of the $\alpha$ scattering on $^{48}$Ti at 240~MeV. 
Here, (a) and (b) show the results in the logarithmic
and linear scales, respectively.
As can be recognized in Fig.~\ref{dcs-alpha.fig}~(b),
the density of C-type-3 best reproduces the experimental
data~\cite{Tokimoto2006}.
Figure~\ref{ampdiff-a.fig} draws the absolute difference
in $g$ between the S-type and C-type-3 configurations at around
the first peak position of the $\alpha$-$^{48}$Ti
elastic scattering cross sections ($\theta=7.2^\circ$).
No difference appears at $b\lesssim 4$ fm because the $\alpha$-nucleus
scattering is strongly absorptive.
The difference is peaked at $b\approx 5$~fm,
which corresponds to the sum of the matter radii of $\alpha$ and $^{48}$Ti.
The $\alpha$-nucleus scattering has no sensitivity to the inner region
and is only sensitive to the outer region of the nuclear density,
while the proton-nucleus scattering probes the density profile
near the nuclear surface.
Thus, it could be interpreted that while the surface region of $^{48}$Ti
is explained by the $jj$-coupling shell model configuration,
the tail region is better explained by the $\alpha$ clustering configuration. 
This result suggests the possibility of a change in structure
as a function of the distance from the center,
from the $jj$-coupling shell model to the cluster model.
We remark that a similar phenomenon has been discussed in $^{44}$Ti, where the $\alpha$ cluster structure is completely broken in the region at small $\alpha$--$^{40}$Ca distances due to the strong spin-orbit contribution. However, with increasing relative distances, the $\alpha$ cluster structure appears beyond the interaction range of the spin-orbit interaction from the $^{40}$Ca nucleus.
Here, the tensor interaction plays a crucial role
in the $\alpha$ clustering~\cite{Ishizuka22}.

\section{Conclusion}
\label{conclusion.sec}
\par
The $\alpha$ clustering in medium-mass nuclear systems is currently a topic of much discussion, and recent analysis of the $\alpha$ knockout reaction on $^{48}$Ti has raised questions about whether $^{48}$Ti is shell-like or cluster-like.
To address this issue, this study has been conducted that involves calculating the proton- and $\alpha$-$^{48}$Ti elastic scattering.
Four types of density distributions were generated,
including the $jj$-coupling shell model and three cluster model configurations, fully microscopically with AQCM.
The Glauber model was used to obtain these cross sections.
We found that the $jj$-coupling shell model configuration best reproduces
the experimental value of the high-energy
proton-nucleus elastic scattering cross sections at the first and second diffraction peaks, which are sensitive to the surface region of the wave function.

On the other hand, a comparison of theoretical and experimental
cross sections of the $\alpha$-nucleus elastic scattering clarifies
the importance of the $\alpha$ clustering in the tail region.
These results suggest that the structure of the nucleus changes as a function of the distance from the center. The $jj$-coupling shell model structure dominates the surface region of the nuclear system, but the structure changes to an $\alpha$+$^{44}$Ca cluster structure in the tail region, in agreement with the analysis of the $\alpha$ knockout reaction.

The study shows that
although the $jj$-coupling shell model wave function dominates around the surface region of $^{48}$Ti,
$\alpha$ clustering is important in the tail region of the wave function.
Understanding such a structural change in the tail region could provide an explanation for the clustering beyond medium-mass nuclei, leading to a more comprehensive understanding of $\alpha$ decay. 
Traditional shell and mean-field models significantly underestimate the $\alpha$ decay probabilities of heavy nuclei, which could be 
improved by incorporating the current mechanism.

\acknowledgments
This work was in part supported by JSPS KAKENHI Grants
Nos.\ 18K03635, 22H01214, and 22K03618.


\begin{thebibliography}{99}
\bibitem{Mayer}
M.~G. Mayer and H.~G. Jensen,
``Elementary theory of nuclear shell structure", John Wiley, Sons, New York, Chapman, Hall, London, (1955).
\bibitem{Michel86}
  F. Michel, G. Reidemeister, and S. Ohkubo, Phys. Rev. Lett.
 {\bf 57}, 1215 (1986).
\bibitem{Yamaya90} T. Yamaya, S. Oh-ami, M. Fujiwara, T. Itahashi, K. Katori, M.
  Tosaki, S. Kato, S. Hatori, and S. Ohkubo, Phys. Rev. C {\bf 42},
1935 (1990).
\bibitem{Yamaya98} T. Yamaya, K. Katori, M. Fujiwara, S. Kato, and S. Ohkubo,
Prog. Theor. Phys. Suppl. {\bf 132}, 73 (1998).
\bibitem{Bailey19}
S.~Bailey, T.~Kokalova, M.~Freer, C. Wheldon, R. Smith, 
J. Walshe, N. Curtis, N. Soi\'c, L. Prepolec, V. Toki\'c {\it et al.}, 
Phys. Rev. C {\bf 100}, 051302(R) (2019).
\bibitem{Taniguchi21}
 Y. Taniguchi, K. Yoshida,  Y. Chiba, Y. Kanada-En'yo, M. Kimura, and K. Ogata,
  Phys. Rev. C {\bf 103}, L031305 (2021).
 \bibitem{AQCM01}
  N. Itagaki, H. Masui, M. Ito, and S. Aoyama, Phys. Rev. C {\bf 71},
064307 (2005).
\bibitem{AQCM02} H. Masui and N. Itagaki, Phys. Rev. C {\bf 75}, 054309 (2007).
\bibitem{AQCM03}T. Yoshida, N. Itagaki, and T. Otsuka, Phys. Rev. C {\bf 79}, 034308 (2009).
\bibitem{AQCM04} N. Itagaki, J. Cseh, and M. P{\l}oszajczak, Phys. Rev. C {\bf 83},
014302 (2011).
\bibitem{AQCM05} T. Suhara, N. Itagaki, J. Cseh, and M. P{\l}oszajczak, Phys. Rev.
  C {\bf 87}, 054334 (2013).
\bibitem{AQCM06} N. Itagaki, H. Matsuno, and T. Suhara, Prog. Theor. Exp. Phys.
  {\bf 2016}, 093D01 (2016).
\bibitem{AQCM07} H. Matsuno, N. Itagaki, T. Ichikawa, Y. Yoshida, and Y.
Kanada-En'yo, Prog. Theor. Exp. Phys. {\bf 2017}, 063D01 (2017).
\bibitem{AQCM08} H. Matsuno and N. Itagaki, Prog. Theor. Exp. Phys. {\bf 2017},
123D05 (2017).
\bibitem{AQCM09} N. Itagaki, Phys. Rev. C {\bf 94}, 064324 (2016).
\bibitem{AQCM10} N. Itagaki and A. Tohsaki, Phys. Rev. C {\bf 97}, 014307 (2018).
\bibitem{AQCM11} N. Itagaki, H. Matsuno, and A. Tohsaki, Phys. Rev. C {\bf 98},
044306 (2018).
\bibitem{AQCM12} N. Itagaki, A. V. Afanasjev, and D. Ray, Phys. Rev. C {\bf 101}, 034304 (2020).
\bibitem{AQCM13} N. Itagaki, T. Fukui, J. Tanaka, and Y. Kikuchi, Phys. Rev. C
  {\bf 102}, 024332 (2020).
\bibitem{AQCM14} N. Itagaki and T. Naito, Phys. Rev. C {\bf 103}, 044303 (2021).
\bibitem{AQCM-Glauber-1} W. Horiuchi and N. Itagaki, Phys. Rev. C {\bf 106}, 044330 (2022). 
\bibitem{AQCM-Glauber-2} W. Horiuchi and N. Itagaki, Phys. Rev. C {\bf 107}, L021304 (2023). 
\bibitem{AQCM-Glauber-3} Y. Yamaguchi, W. Horiuchi, and N. Itagaki,  Phys. Rev. C {\bf 108}, 014322 (2023).
\bibitem{Brink}
 D.~M. Brink, Many-body description of nuclear structure and reactions, in 
 {\it Proceedings of the International  School of Physics ``Enrico Fermi'', Course XXXVI}, edited by L. Bloch (Achademic Press New York, 1966), p. 247.
\bibitem{Horiuchi07} W. Horiuchi, Y. Suzuki, B. Abu-Ibrahim, and A. Kohama, Phys. Rev. C {\bf 75}, 044607 (2007); {\it ibid} {\bf 76}, 039903(E) (2007).
\bibitem{Horiuchi12} W. Horiuchi, T. Inakura, T. Nakatsukasa, and Y. Suzuki,
         Phys. Rev. C {\bf 86}, 024614 (2012).
\bibitem{Terashima08} S. Terashima, H. Sakaguchi, H. Takeda, T. Ishikawa, M. Itoh, T. Kawabata, T. Murakami, M. Uchida, Y. Yasuda, M. Yosoi {\it et al.},
         Phys. Rev. C {\bf 77}, 024317 (2008).
       \bibitem{Zenihiro10} J. Zenihiro, H. Sakaguchi, T. Murakami, M. Yosoi, Y. Yasuda, S. Terashima, Y. Iwao, H. Takeda, M. Itoh, H.~P. Yoshida,
         and M. Uchida, Phys. Rev. C {\bf 82}, 044611 (2010).
\bibitem{Hatakeyama18} S. Hatakeyama, W. Horiuchi, and A. Kohama, Phys. Rev. C {\bf 97}, 054607 (2018).
\bibitem{Glauber} R.~J. Glauber, {\it Lectures in Theoretical Physics}, edited by W.~E. Brittin and L.~G. Dunham (Interscience, New York, 1959), Vol. 1, p.315.
\bibitem{Suzuki03} 
 Y. Suzuki, R.~G. Lovas, K. Yabana, K. Varga,
 {\it Structure and reactions of light exotic nuclei}
 (Taylor \& Francis, London, 2003).
\bibitem{Ibrahim08} B. Abu-Ibrahim, W. Horiuchi, A. Kohama, and Y. Suzuki, Phys. Rev. C {\bf 77}, 034607 (2008); {\it ibid} {\bf 80}, 029903(E) (2009); {\bf 81}, 019901(E) (2010).
\bibitem{Horiuchi16} W. Horiuchi, S. Hatakeyama, S. Ebata, and Y. Suzuki, Phys. Rev. C {\bf 93}, 044611 (2016).  
 \bibitem{Hatakeyama19} S. Hatakeyama and W. Horiuchi, Nucl. Phys. {\bf A 985}, 20 (2019).  
 \bibitem{Varga02} K. Varga, S.~C. Pieper, Y. Suzuki, and R.~B. Wiringa, Phys. Rev. C {\bf 66}, 034611 (2002).
\bibitem{Ibrahim09} B. Abu-Ibrahim, S. Iwasaki, W. Horiuchi, A. Kohama, and Y. Suzuki, J. Phys. Soc. Jpn., Vol. 78, 044201 (2009).
\bibitem{Nagahisa18} T. Nagahisa and W. Horiuchi, Phys. Rev. C {\bf 97}, 054614 (2018).
 \bibitem{Angeli13} I. Angeli and K.~P. Marinova,
   At. Data Nucl. Data Tables {\bf 99}, 69 (2013).  
\bibitem{Alkhazov76} G.~D. Alkhazov, T. Bauer, R. Beurtey, A. Boudard, G. Bruge,
A. Chaumeaux, P. Couvert, G. Cvijanovich, H.~H.~Duhm, J.~M. Fontaine {\it et al.}, 
Nucl. Phys. {\bf A 274}, 443 (1976).
\bibitem{Comparat1974}
V.~Comparat, R.~Frascaria, N.~Marty, M.~Morlet, and A.~Willis, 
Nucl. Phys. {\bf 221}, 403 (1974).
\bibitem{Tokimoto2006}
Y. Tokimoto, Y.-W. Lui, H. L. Clark, B. John, X. Chen, and D. H. Youngblood,
Phys. Rev. C {\bf 74}, 044308 (2006).
 \bibitem{Hofstadter56} R. Hofstadter, Rev. Mod. Phys. {\bf 28}, 214 (1956).  
\bibitem{Kamimura81}
M. Kamimura, Nucl. Phys. {\bf A351}, 456 (1981).
\bibitem{FF1987}
H. De Vries, C.~W. De Jager, and C. De Vries,
Atom. Data Nucl. Data Tabl. {\bf 36}, 495 (1987).
\bibitem{FF1974}
C.~W. De Jager, H. De Vries,  and C. De Vries,
Atom. Data Nucl. Data Tabl. {\bf 14}, 479 (1974).
\bibitem{Ishizuka22} C.~Ishizuka, H.~Takemoto, Y.~Chiba,
A.~Ono, and N.~Itagaki, Phys. Rev. C {\bf 105}, 064314 (2022).
\end{thebibliography}
\end{document}